\documentclass[aps,superscriptaddress,amsmath,amssymb,floatfix,twocolumn,showpacs,amsfonts,longbibliography]{revtex4-1}
\usepackage{times}
\usepackage[varg]{txfonts}
\usepackage{textcomp}
\usepackage{graphicx}
\usepackage{subfigure}
\usepackage{tabu}
\usepackage{color}
\usepackage[colorlinks=true,citecolor=blue,urlcolor=blue,linkcolor=blue,hyperindex]{hyperref}
\usepackage{braket}
\usepackage{float}
\usepackage{overpic}
\usepackage{amssymb}

\allowdisplaybreaks

\usepackage{times}

\def\be{\begin{equation}} \def\ee{\end{equation}}
\def\bea{\begin{eqnarray}} \def\eea{\end{eqnarray}}

\def\bk{{\bf k}}

\def\be{{\bf e}}

\begin{document}

\title{Unconventional Topological Insulators from Extended Topological Band Degeneracies}

\author{Zhongbo Yan} \email{yanzhb5@mail.sysu.edu.cn}
\affiliation{School of Physics, Sun Yat-Sen University, Guangzhou 510275, China}

\begin{abstract}
A general and beautiful picture for the realization of topological insulators is
that the mass term of the Dirac model has a nodal surface wrapping one Dirac point. We
show that this geometric picture based on Dirac points can be generalized to extended band degeneracies
with nontrivial topological charges. As the nontrivial topological charges force
the extended band degeneracies to be created or annihilated in pairs, 
when the nodal surface of mass wraps one such extended band degeneracy, the resulting gapped phase 
must be topologically nontrivial since it cannot adiabatically be deformed 
into a topologically trivial atomic insulator without closing the energy gap. We use
nodal lines carrying a nontrivial $Z_{2}$ monopole charge in three dimensions to illustrate the physics.
Notably, because the wrapping surfaces for an extended band degeneracy
are diverse, we find that this generalization can bring topological insulators
with unconventional pattern of boundary states. 
\end{abstract}

\maketitle

Dirac models paly a fundamental role in the description
of topological insulators and superconductors\cite{hasan2010,Qi2011review,ryu2010,
shen2013topological,bernevig2013topological}. In $D$ dimensions,
their tight-binding forms are simply given by
\begin{eqnarray}
H(\bk)=\sum_{i=1}^{D}\lambda_{i}\sin k_{i}\Gamma_{i}+m(\bk)\Gamma_{D+1},\label{Dirac}
\end{eqnarray}
with the matrices $\Gamma_{1,2,...,D+1}$ satisfying the Clifford algebra, i.e., $\{\Gamma_{i},\Gamma_{j}\}=2\delta_{ij}$.
$m(\bk)$ is an even function of the momentum and known as the Dirac mass. When $m(\bk)$ vanishes identically,
the energy bands touch and form Dirac points at time-reversal invariant momenta. A universal rule
for nonaccidental Dirac points is that they must be created or annihilated in pairs, 
known as the fermion-doubling theorem\cite{Nielsen1981nogo}.
When the existence of some symmetries forces the absence of additional mass terms which anti-commute
with all terms in Eq.(\ref{Dirac}), i.e., the Dirac
mass term  $m(\bk)\Gamma_{D+1}$ is unique due to symmetry constraint,  a strong topological insulator
or superconductor is realized when the zeros of $m(\bk)$, dubbed {\it mass nodal surface} (MNS) in this paper,
wraps only one of the Dirac points. The underlying mechanism is simply because the resultant gapped 
phase for such a configuration cannot continuously be  deformed into the topologically trivial atomic limit 
($\lambda_{i}=0$ and $m(\bk)=m$ with $m$ a nonzero constant) without closing the energy gap.

Band degeneracies, however, are not restricted to the form of point.
They can also be extended in the momentum space and form very rich
structures. For instance, one-dimensional band degeneracy, also known
as nodal line,  can display
structures of the form from rings\cite{Burkov2011nodal,Yu2015,Kim2015,Fang2015nodal} and chains\cite{Bzdusek2016chain,Yu2017chain,Yan2018chain,Gong2018chain}
to exotic links\cite{Chen2017link,Yan2017link,Ezawa2017link,Chang2017link,Yang2019link} and knots\cite{Bi2017knot}.
Over the past few years, materials with such one-dimensional band degeneracies near the Fermi energy
have attracted great interest\cite{fang2016nlsm,Yang2018review,Klemenz2019review}. Similarly to the Dirac point, an extended band degeneracy requires
appropriate crystal symmetries to protect its stability. By breaking the protecting symmetry,
it may be split  into some lower-dimensional band degeneracies
or completely gapped out. With the Dirac models in mind, the latter situation
attracts us to investigate the underlying topological property of the resultant gapped
phase for which the mass term has a nodal surface
wrapping only one extended band degeneracy.

Extended band degeneracies, unlike Dirac points, are in principle unnecessary to obey the fermion-doubling theorem.
As a result, we need to divide them into two classes. For the first class, they do not carry
a global topological charge and can be singly created or annihilated by tuning the Hamiltonian parameters
without breaking the protecting symmetries\cite{Fang2015nodal}. For the second class, they carry a nontrivial global topological charge
and therefore must be created or annihilated in pairs like the Dirac points\cite{Fang2015nodal,Bzdusek2017charge,Ahn2018charge,Song2018charge}. For the first class, it is obvious that
the resultant gapped phase must be topologically trivial because we can first deform the extended band degeneracy to vanish
and then deform the wrapping MNS to vanish without closing the bulk gap. In other words, the insulator
is adiabatically connected to a topologically trivial atomic insulator. For the second class, it is obvious
that the adiabatic path to the atomic insulator is obstructed, just like the scenario
in  conventional topological insulators, therefore, the resultant gapped phase must be topologically nontrivial.
Because of the distinction in dimensions for the wrapped band degeneracy,
it is natural to expect that this new scenario may bring
new types of topological phases with unconventional pattern of boundary states.
In this work, we use one-dimensional
nodal lines carrying a nontrivial $Z_{2}$ monopole charge in three dimensions as
a concrete example to demonstrate the above general arguments.

{\it Nodal lines protected by a $Z_{2}$ monopole charge---} Similarly to the Dirac model in Eq.(\ref{Dirac}), we divide
the Hamiltonian into two parts, i.e., $H=H_{\rm NL}+H_{\rm M}$, with $H_{\rm NL}$ describing the part
realizing topological band degeneracies and  $H_{\rm M}$ the mass term gapping out the band degeneracies. $H_{\rm NL}$ and 
$H_{\rm M}$ satisfy $\{H_{\rm NL},H_{\rm M}\}=0$.
In Ref.\cite{Fang2015nodal}, Fang {\it et al} showed that when spin-orbit coupling is negligible and a combination of
inversion symmetry (P) and time-reversal symmetry (T) is present, nodal lines carrying a nontrivial $Z_{2}$ monopole charge
can be realized in a four-band minimal model. Here we consider a tight-binding generalization of the model in
Ref.\cite{Fang2015nodal}, which reads
\begin{eqnarray}
H_{\rm NL}(\bk)&=&\lambda_{z} \sin k_{z}s_{z}+\lambda_{x} \sin k_{x}s_{x}+\lambda_{y} \sin k_{y}\sigma_{y}s_{y}\nonumber\\
&&+(t_{0}-t\cos k_{x}-t\cos k_{y}-t\cos k_{z})\sigma_{x}s_{x},\label{Hamiltonian}
\end{eqnarray}
where $\sigma_{i}$ and $s_{i}$ act on certain pseudo-spin degrees of freedom. For notational simplicity,
the lattice constant is set to unit throughout this work. This Hamiltonian has
both $P$ and $T$, with $P=\sigma_{z}s_{y}$ and $T=\sigma_{z}s_{y}K$ ($K$ denotes the complex conjugation). 
Since $[P,T]=0$ and $(PT)^{2}=1$, the energy bands of
the Hamiltonian do not have the double degeneracy as in a spinful system with both $P$ and $T$. However,
here $T^{2}$ also gives $-1$, so Kramers degeneracy are still present at time-reversal invariant momenta.

\begin{figure}
\subfigure{\includegraphics[width=8cm, height=6cm]{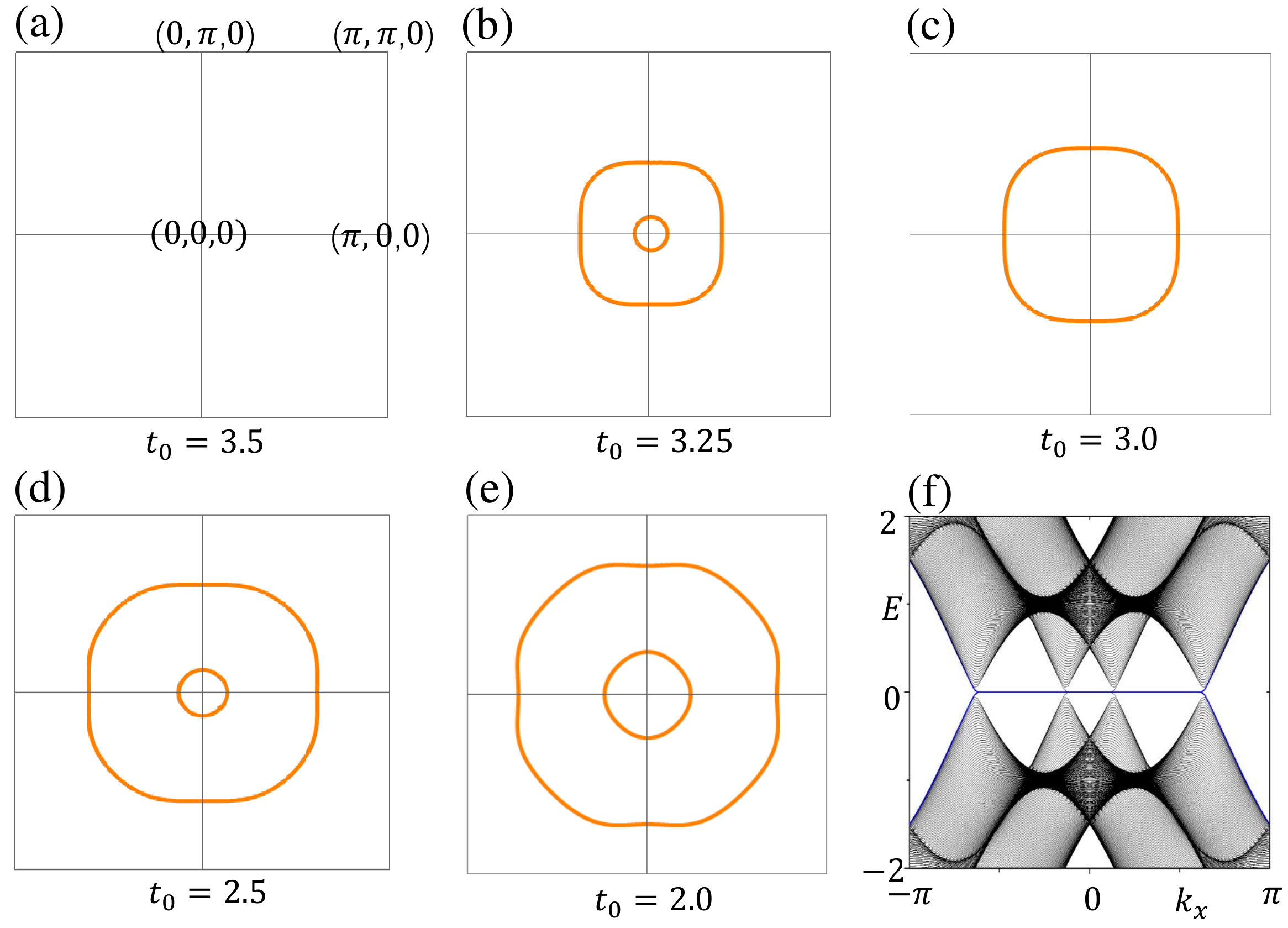}}
\caption{ (a)-(e) The evolution of $Z_{2}$-monopole-charge nodal lines in the $k_{z}=0$ plane. (a) no nodal lines exist,
(b) two concentric nodal lines emerge, (c) one of the nodal lines shrinks to a point, (d) the point re-expands to
a line, (e) the size of nodal lines increases further. (f) Energy spectra for a geometry with open boundary condition
in the $z$ direction and periodic boundary condition in
both $x$ and $y$ directions. $k_{y}=0$, $t_{0}=2.5$, and the system size in the $z$ direction is $L_{z}=100$.
Characteristic drumhead surface states appear within
the projections of the nodal lines.    Common parameters: $\lambda_{x,y,z}=t=1$.
}  \label{configuration}
\end{figure}

The energy spectra for $H_{\rm NL}(\bk)$ read
\begin{eqnarray}
E(\bk)=\pm\sqrt{(\lambda_{z}\sin k_{z})^{2}+(A(\bk)\pm B(\bk))^{2}},
\end{eqnarray}
where $A(\bk)=\sqrt{(\lambda_{x}\sin k_{x})^{2}+(\lambda_{y}\sin k_{y})^{2}}$
and $B(\bk)=(t_{0}-t\cos k_{x}-t\cos k_{y}-t\cos k_{z})$. The nodal lines are located
at either the $k_{z}=0$ plane or the $k_{z}=\pi$ plane. At the $k_{x}=0$ plane, they
are determined by $\sqrt{(\lambda_{x}\sin k_{x})^{2}+(\lambda_{y}\sin k_{y})^{2}}=\pm
(t_{0}-t-t\cos k_{x}-t\cos k_{y})$. At the $k_{x}=\pi$ plane, they
are determined by $\sqrt{(\lambda_{x}\sin k_{x})^{2}+(\lambda_{y}\sin k_{y})^{2}}=\pm
(t_{0}+t-t\cos k_{x}-t\cos k_{y})$. Without losing generality, in this work
we consider the simplest situation with only two concentric nodal lines located
at the $k_{z}=0$ plane, as shown in Fig.\ref{configuration}. From the evolution of nodal lines 
presented in Figs.\ref{configuration}(a)-(e), one can infer the fact that
for this Hamiltonian,  a nodal line enclosing a time-reversal invariant momentum cannot shrink to a point and vanish.
In Ref.\cite{Fang2015nodal}, Fang {\it et al} explained that the underlying reason is because the nodal line
carries a nontrivial $Z_{2}$ monopole charge. Without explicitly calculating this topological invariant in terms of the wave functions of occupied states, it in fact can be intuitively understood
by noting the fact that when the nodal line shrinks to a point at a time-reversal invariant momentum,
it reduces to a Dirac point at this critical situation (see Eq.(\ref{Hamiltonian})), suggesting that
the $Z_{2}$ monopole charge of the nodal line is inherited from the Dirac point.
It is noteworthy that while  nodal lines carrying nontrivial and trivial $Z_{2}$ monopole charges display remarkable
distinction in stability,  the associated surface states, however, take the same
characteristic drumhead form, as shown in Fig.\ref{configuration}(f).

{\it MNS of the sphere structure---} Let us first investigate the situation that
the wrapping MNS is of the sphere structure. To be specific, we let
\begin{eqnarray}
H_{\rm M}(\bk)=(m_{0}-\cos k_{x}-\cos k_{y}-\cos k_{z})\sigma_{x}s_{y}.
\end{eqnarray}
It is readily checked that the Pauli matrix form of 
$H_{\rm M}$ is the only possibility that anti-commutes with all terms in $H_{\rm NL}$.
One can also check that $[T, H_{\rm M}]=0$ but $[P, H_{\rm M}]\neq 0$, suggesting that the introduction of $H_{\rm M}$
does not break the time-reversal symmetry, but breaks the inversion symmetry and their combinational symmetry (the $PT$ symmetry
which forces the Hamiltonian to be real), consequently gapping out the nodal lines.

\begin{figure}
\subfigure{\includegraphics[width=8cm, height=8cm]{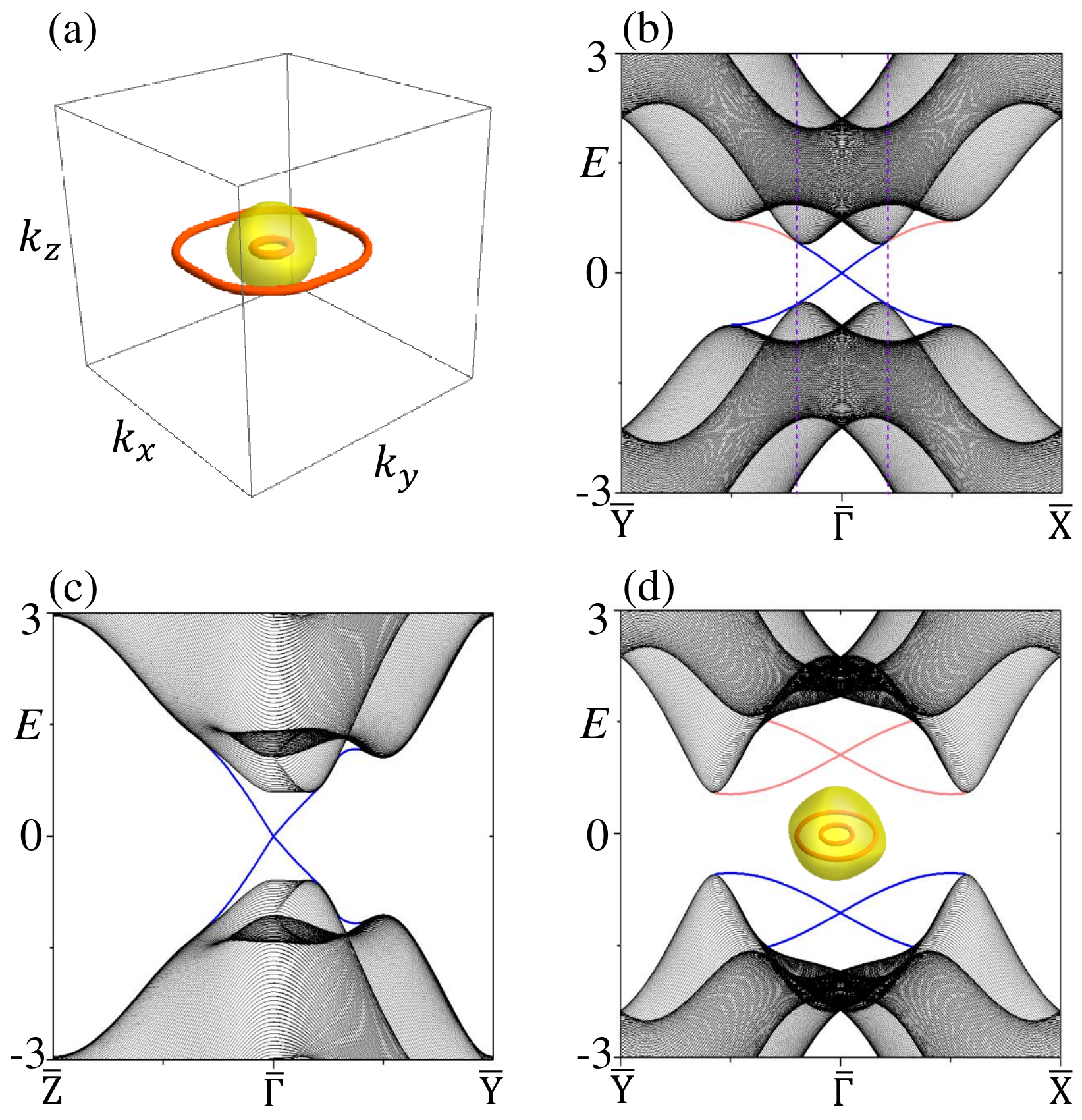}}
\caption{ (a) A sphere-form MNS (yellow) wraps the nodal line (red)
with smaller size. (b)  Between the two dashed purple lines, the in-gap dispersions are doubly degenerate,
corresponding to the existence of a single Dirac cone on each $z$-normal surface. Outside, the blue and
red lines are without degeneracy and represent boundary states on the lower and upper surface, respectively.
(c) Each $x$-normal or $y$-normal surface also harbors a single Dirac cone, here we have only presented
the results for $x$-normal surfaces because the system has rotational symmetry in the $xy$ plane.
(d) When the MNS wraps both nodal lines (the inset),
the configuration realizes a trivial insulator but with surface Dirac cones coexisting with bulk states
on the $z$-normal surfaces. The parameters are: $\lambda_{x,y,z}=t=1$ and $t_{0}=m_{0}=2.5$ in (a)(b)(c);
$\lambda_{z}=t=1$, $\lambda_{x}=\lambda_{y}=0.5$,  $t_{0}=2.5$, and $m_{0}=1.5$ in (d); $L_{z}=100$ in (b)(d),
and $L_{x}=100$ in (c).
}  \label{sphere}
\end{figure}

Let us consider that the MNS determined by $\sum_{i=x,y,z}\cos k_{i}=m_{0}$ wraps
the nodal line with smaller size. For concreteness, we let $m_{0}=t_{0}=2.5$, which naturally satisfies
the requirement, as shown in Fig.\ref{sphere}(a). Apparently, because the two nodal lines can only be annihilated
in pairs, no matter how we deform the MNS, inwardly or outwardly,
it will inevitably cross one of the nodal lines before it gets vanished. 
In other words, the MNS cannot vanish without closing the energy gap.  This
simple fact indicates that the adiabatic path to a topologically trivial atomic insulator
is obstructed, so the configuration shown in Fig.\ref{sphere}(a) must
correspond to a topological insulator. By investigating the boundary states,
we find that each open surface indeed harbors a single Dirac cone, as shown in
Figs.\ref{sphere}(b)(c). Because of $T^{2}=-1$ as mentioned previously, the Dirac cones
are centered at the time-reversal invariant momentum $\bar{\Gamma}$ of the surface
Brillouin zone. Just like in a spinful topological
insulator, here the Dirac cones are also robust against
perturbations respecting time-reversal symmetry. These results demonstrate 
the realization of a topological insulator with robust boundary states. 
For comparison, we have also investigated the configuration with
the MNS wrapping both nodal lines (see the inset of
Fig.\ref{sphere}(d)). Owing to the $Z_{2}$ nature of the nodal lines, we find 
that as expected there is no boundary states connecting
the conduction and valence bands, demonstrating that the configuration corresponds to a topologically
trivial insulator. Nevertheless, we find that Dirac cones  still appear on the $z$-normal surfaces, though
their energy are shifted away from the gap, as shown in Fig.\ref{sphere}(d).

\begin{figure}
\subfigure{\includegraphics[width=8cm, height=8cm]{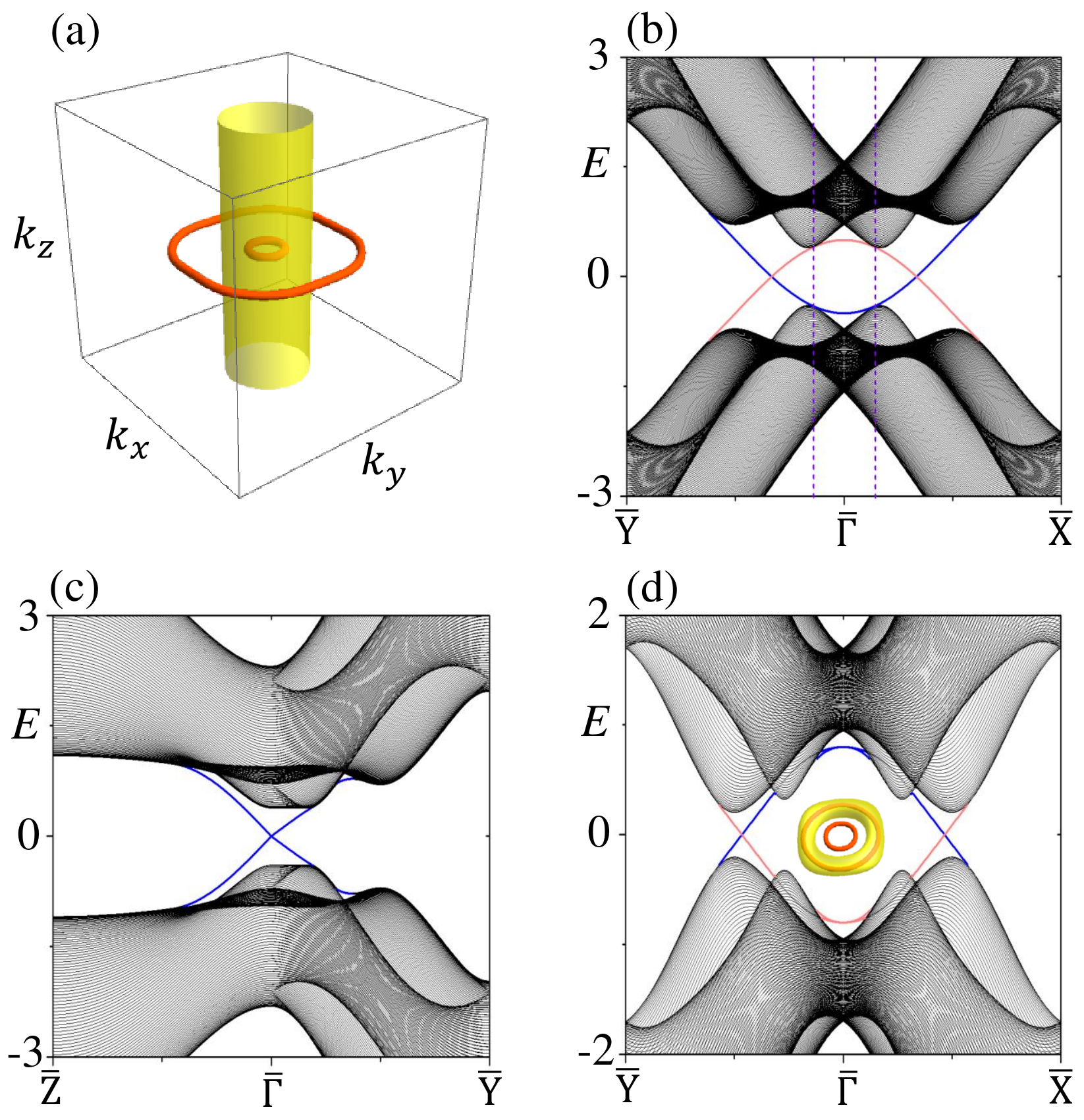}}
\caption{ (a) A torus-form MNS (yellow) wraps the nodal line (red)
with smaller size. (b) Between the two dashed purple lines, the dispersions in red color
and blue color are doubly degenerate, which correspond to trivial boundary states on
the upper and lower $z$-normal surface, respectively. Outside,
the in-gap dispersions are without degeneracy and correspond to topological boundary states.
(c) Each $x$-normal surface (also for $y$-normal surfaces) harbors a single Dirac cone. (d) Boundary
states on the $z$-normal surfaces for the configuration that the torus-form
MNS encloses the nodal line with larger size (the inset).
The parameters are:  $\lambda_{x,y,z}=t=1$, $t_{0}=2.5$ and $m_{0}=1.5$ in (a)(b)(c);
$\lambda_{z}=t=1$, $\lambda_{x}=\lambda_{y}=0.5$, $t_{0}=2.5$, $B_{0}=2.0$ and
$\delta=0.2$ in (d); $L_{z}=100$ in (b)(d), and $L_{x}=100$ in (c).
}  \label{cylinder}
\end{figure}

{\it MNS of the torus structure---} For a nodal line, the wrapping
MNS can also take the torus structure. For simplicity, we first consider
a mass term of the form
\begin{eqnarray}
H_{\rm M}(\bk)=(m_{0}-\cos k_{x}-\cos k_{y})\sigma_{x}s_{y}.
\end{eqnarray}
Because of the periodicity of the Brillouin zone, $H_{\rm M}$ gives
a torus-form MNS for $-2<m<2$. Let us again consider that
the MNS encloses the nodal line with smaller size, as shown in Fig.\ref{cylinder}(a).
It is easy to see that this configuration is adiabatically connected to the configuration in
Fig.\ref{sphere}(a), so it must also realize a topological insulator.
Notably, we find that the boundary states on the $z$-normal surfaces for this configuration
no longer take the Dirac cone structure, instead they take a bowl structure, as shown in
Fig.\ref{cylinder}(b). Anyway, the boundary states still connect the conduction and valence bands,
confirming the nontrivial band topology. To the best of our knowledge,
such bowl-like surface states have previously only been found
in magnetic Hopf insulators\cite{Moore2008hopf,Yan2017link,Alexandradinata2019hopf}. 
This coincidence has a natural explanation.
In the Hopf insulators, the bowl-like surface states are tied to
the underlying link structure. In Fig.\ref{cylinder}(a), the nodal line
with larger size and the torus-form MNS in fact
form an extra link structure similar to that in the Hopf insulator.
In spite of similarity, we need to emphasize that there also
exist remarkable difference between them. In the Hopf insulators,
a bowl-like surface state is of one-band nature and spin-polarized due 
to the absence of time-reversal symmetry\cite{Moore2008hopf,Yan2017link,Alexandradinata2019hopf}. 
In Fig.\ref{cylinder}(b), while the bowl-like surface state on each surface
 also looks like of one-band nature, 
the time-reversal symmetry forces the pseudo-spin and the momentum to be locked. 
That is, the bowl-like surface states in Fig.\ref{cylinder}(b) exhibit nontrivial pseudo-spin
textures just like the Dirac surface states. An intuitive understanding of the bowl structure and 
the nontrivial pseudo-spin textures is that each bowl-like surface state in the gap corresponds to 
one part of a Dirac cone, with the other part containing the Dirac point buried in the bulk states. 

For $x$-normal and $y$-normal surfaces, we find that each surface still harbors
a single Dirac cone, as shown in Fig.\ref{cylinder}(c). The coexistence of surface Dirac cones and
bowl-like topological surface states reveals a new type of topological insulating phases unexplored before.
Notably, the total Hamiltonian $H=H_{\rm NL}+H_{\rm M}$ still only involves at most nearest-neighbor hoppings,
its simplicity holds great promise for the realization of this type of topological insulators, both in
real materials and in artificial systems.

While from the above analysis it is apparent that a topological insulator should also be realized
when the MNS encloses only the nodal line with larger size,
we demonstrate this explicitly for completeness. To be specific, we consider
a mass term of the form
\begin{eqnarray}
H_{\rm M}(\bk)=[\sin^{2}k_{z}+(B_{0}-\sum_{i=x,y,z}\cos k_{i})^{2}-\delta]\sigma_{x}s_{y},
\end{eqnarray}
where $B_{0}$ and $\delta$ are two real parameters controlling the position and the size of the
MNS. Fig.\ref{cylinder}(d) shows that the boundary states on the $z$-normal surfaces
also take the bowl structure when the MNS forms a torus enclosing the nodal line
with larger size (see the inset). The result can be simply understood by noting
the torus in Fig.\ref{cylinder}(a) can be continuously deformed to the torus in
Fig.\ref{cylinder}(d). In addition, one can find from Fig.\ref{cylinder}(d) that
the cross section of the MNS for a given toroidal angle links with
the nodal line.

\begin{figure}
\subfigure{\includegraphics[width=8cm, height=4cm]{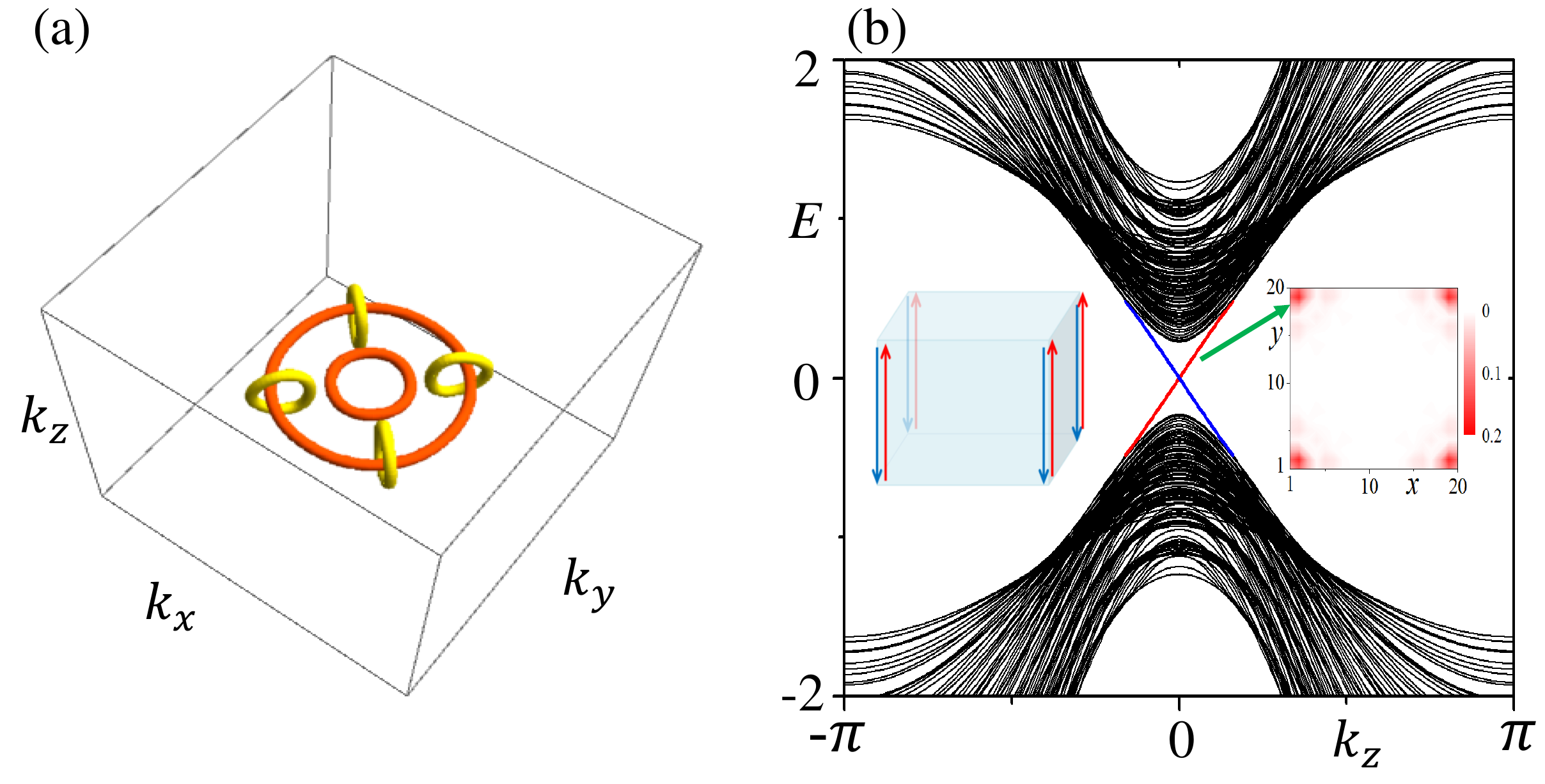}}
\caption{ (a) Nodal lines from $H_{\rm NL}$ and $H_{\rm M}$ form link structure.
(b) The in-gap dispersions in red color (up-moving states) and blue color (down-moving states)
are of four-fold degeneracy, corresponding to four pairs of gapless helical states localized
on the four hinges. The left inset is a schematic diagram for the distribution of the gapless helical states, and
the right inset shows the probability density profiles of the four branches of upmoving states at $k_{z}=0.1$.
The parameters are: $\lambda_{z}=t=1$, $\lambda_{x}=\lambda_{y}=0.5$, $t_{0}=2.5$, $B_{0}=2.0$,
$\delta=0.3$, and $L_{x}=L_{y}=20$.
}  \label{generalization}
\end{figure}

{\it Generalization---} It is apparent that the geometry picture exemplified in three dimensions is also 
applied to dimensions higher than three. While the spatial dimension of real materials
is bounded by three, topological insulators in dimensions $D\geq4$ can be simulated in terms 
of synthetic dimensions which are realized by certain controllable 
internal degrees of freedom or periodic parameters\cite{Celi2014synthetic,Price2015FDQHE,
Ozawa2016synthetic,Ozawa2019}. Notably, 
extended band degeneracy become more common and diverse in higher dimensions, which 
consequently allows the formation of more exotic configurations for the MNS and extended 
band degeneracy.

Another interesting direction for generalization is the dimension of the MNS. 
This direction is relevant to systems with more than one symmetry-allowed mass term. As mentioned
previously, such a situation implies a trivial insulator. However, the triviality is to
the first-order level. Very recently, the concept dubbed higher-order topological insulator
indicates that many classes of insulators which were believed to be trivial
can be topologically nontrivial
in a higher-order sense\cite{benalcazar2017quantized,Schindler2018HOTIa,Sitte2012,Zhang2013surface, Benalcazar2017prb,Song2017higher,Langbehn2017hosc}. That is, topological gapless states can appear at some lower-dimensional
boundaries even when the one-dimensional lower boundaries do not harbor topological gapless states.

It is readily seen that when $H_{M}$ contains at least two anti-commuting
terms and $\{H_{\rm NL}, H_{\rm M}\}=0$ is still hold, the MNS
becomes the band degeneracy of $H_{\rm M}$ and its dimension will be reduced.
The dimension reduction of the MNS, however, does not rule out the possibility for
the MNS and extended band degeneracy to form configurations that are not adiabatically connected to
the topologically trivial atomic limit.  For instance, when two mass terms are allowed by symmetry in three
dimensions, the MNSs also become lines which can form topological link structures
with the nodal lines from $H_{\rm NL}$.  As an example,
we consider the following Hamiltonian,
\begin{eqnarray}
H(\bk)&=&H_{\rm NL}\otimes \tau_{z}+[\sin^{2}k_{z}+(B_{0}-\sum_{i=x,y,z}\cos k_{i})^{2}-\delta]\sigma_{x}s_{y}\tau_{z}\nonumber\\
&&+(\cos k_{x}-\cos k_{y})\tau_{x},
\end{eqnarray}
where $\tau_{i}$ act on another pseudospin. While each nodal line is doubled due
to the introduction of a new pseudospin, they
cannot be annihilated with their degenerate partners if the pseudospin is conserved.
Because of the doubling, it is readily checked
that the last two mass terms do not break the time-reversal symmetry and chiral symmetry
of $H_{\rm NL}\otimes \tau_{z}$ (the chiral operator is $\tau_{y}$).  
As shown in Fig.\ref{generalization}(a), for appropriate $B_{0}$ and $\delta$,
the nodal lines originated from the
mass terms form link structure with the nodal lines from $H_{\rm NL}\otimes \tau_{z}$.
By considering a geometry with open boundary condition in the $xy$ plane and periodic boundary
condition in the $z$ direction, we find that gapless helical states appear at the one-dimensional hinges
of the geometry (see Fig.\ref{generalization}(b)), indicating that the link configuration in
Fig.\ref{generalization}(a) realizes
a higher-order topological insulator.

{\it Discussions and Conclusions.---} It is easy to see that the decomposition $H_{\rm NL}+H_{\rm M}$ can
also be decomposed as $H_{\rm D}+H_{\rm SB}$, where $H_{\rm D}$ takes the Dirac form given in Eq.(\ref{Dirac}) and
$H_{\rm SB}$ denotes symmetry-breaking terms. Therefore, the geometric picture developed in this work
can also be applied to analyze the topological property of insulators without certain symmetries, e.g., insulators without 
inversion symmetry.
As is known, when the inversion symmetry is broken, the topological property of a time-reversal invariant insulator
can no longer be simply determined by the parity at time-reversal invariant momenta\cite{fu2007a}.
The geometric picture developed in this work, however, is hold as long as that decompositions
like $H_{\rm NL}+H_{\rm M}$ with $\{H_{\rm NL}, H_{\rm M}\}=0$ is possible.

In summary, we have developed a new geometric picture for topological insulators
on the basis of extended band degeneracies. Guided by it, we have revealed the existence of new types of
topological insulators with unconventional pattern of boundary states through concrete examples
in three dimensions. The simplicity of the developed
geometric picture allows it to be taken as a guiding principle for the discovery and design
of new types of topological insulators.

This work is supported  by the Startup Grant (No. 74130-18841219) and
the National Science Foundation of China (Grant No.11904417).

\bibliography{dirac}

\end{document}